\documentclass[conference,9pt]{IEEEtran}
\usepackage{lipsum}
\usepackage{float}
\usepackage{footmisc}
\usepackage{cite}
\usepackage{amsmath,amssymb,amsfonts}
\usepackage{algorithmic}
\usepackage{color}
\usepackage{xcolor}
\usepackage{graphicx}
\usepackage{subfigure}
\usepackage{textcomp}
\usepackage{theorem}
\usepackage{multirow,booktabs}
\usepackage[linesnumbered,ruled,vlined]{algorithm2e}
\usepackage{amsmath} 
\UseRawInputEncoding
\usepackage{etoolbox}
\usepackage[T1]{fontenc}
\usepackage[utf8]{inputenc}
\usepackage[normalem]{ulem}
\usepackage{wrapfig}
\usepackage{pifont}

\allowdisplaybreaks

\makeatletter
\patchcmd{\@makecaption}
  {\scshape}
  {}
  {}
  {}
\makeatletter
\patchcmd{\@makecaption}
  {\\}
  {.\ }
  {}
  {}
\makeatother
\def\tablename{Table}

\newcommand{\qadd}[1]{{\color{black}{#1}}}%
\newcommand{\qdel}[1]{}
\newcommand{\todo}[1]{{\color{black}{#1}}}%
\newcommand{\ndel}[1]{}
\newcommand{\nadd}[1]{{\color{black}{#1}}}

\newcommand{\xadd}[1]{{\color{black}{#1}}}
\newcommand{\xdel}[1]{}

\SetKwInput{KwData}{Function}

\newtheorem{Example}{Example}

\renewcommand{\baselinestretch}{0.931}

\newcommand{\MethodName}{MASIM}
\newcommand{\PName}{CPP}
\newcommand{\CombPName}{CICPP}

\title{\MethodName{}: An Efficient Multi-Array Scheduler for In-Memory SIMD Computation}
\author{
    \IEEEauthorblockN{Xingyue Qian$^{1}$, Chen Nie$^{2}$, Zhezhi He$^{2,*}$, and Weikang Qian$^{1,3,*}$}
    \IEEEauthorblockA{$^{1}$University of Michigan-SJTU Joint Institute, $^2$School of Electronic Information and Electrical Engineering, \\and $^3$MoE Key Lab of AI, Shanghai Jiao Tong University, Shanghai, China}
    \IEEEauthorblockA{Emails: \{qianxingyue, chen.nie, zhezhi.he, qianwk\}@sjtu.edu.cn; $^*$corresponding authors}
}
\IEEEoverridecommandlockouts
\begin{document}
\maketitle

\begin{abstract}
Single instruction, multiple data (SIMD) is a popular design style of in-memory computing (IMC) architectures, which enables memory arrays to perform logic operations to achieve low energy consumption and high parallelism.
To implement a target function on the data stored in memory, the function is first transformed into a netlist of the supported logic operations through logic synthesis.
Then, the scheduler transforms the netlist into the instruction sequence given to the architecture.
An instruction is either computing a logic operation in the netlist or copying the data from one array to another.
Most existing schedulers focus on optimizing the execution sequence of the operations to minimize the number of memory rows needed, neglecting the energy-consuming copy instructions, which cannot be avoided when working with arrays with limited sizes.
In this work, our goal is to reduce the number of copy instructions to decrease overall energy consumption.
We propose \MethodName{}, a \underline{m}ulti-\underline{a}rray \underline{s}cheduler for \underline{i}n-\underline{m}emory SIMD computation. It consists of a priority-driven scheduling algorithm \xdel{that uses the number of copy instructions as the primary priority and the change in locality as the secondary priority.}and an iterative improvement process.
Compared to the best state-of-the-art scheduler, \xdel{ours}\MethodName{} reduces the number of copy instructions by 63.2\% on average, which leads to a 28.0\% reduction in energy. 
\end{abstract}

\section{Introduction}\label{sec:intro}
Traditional von Neumann architecture consists of a memory for data storage and a processor for data processing.
It is the foundation of most modern computing systems, but it suffers from a problem called the \emph{memory wall}: the energy spent on data transfer can be orders of magnitude \qdel{more}\qadd{larger} than that on computation~\cite{WALL}.
To break the wall, many \emph{in-memory computing (IMC)} architectures have been proposed to enable memory arrays to perform logic operations such as XOR and majority (MAJ) so that costly data transfer can be reduced~\cite{MAGIC,SIMDRAM,XMG-GPPIC}.
\emph{Single instruction, multiple data (SIMD)} is a prevalent design style of IMC architectures with high parallelism~\cite{28NM,SIMDRAM,XMG-GPPIC}.
In SIMD IMC, the calculation of the target function, \textit{e.g.}, multiplication, on one input pattern is performed within a single column, and different columns are used to calculate the same function on different input patterns in parallel.

\begin{figure}[!htbp]
    \centering
    \includegraphics[scale=0.31]{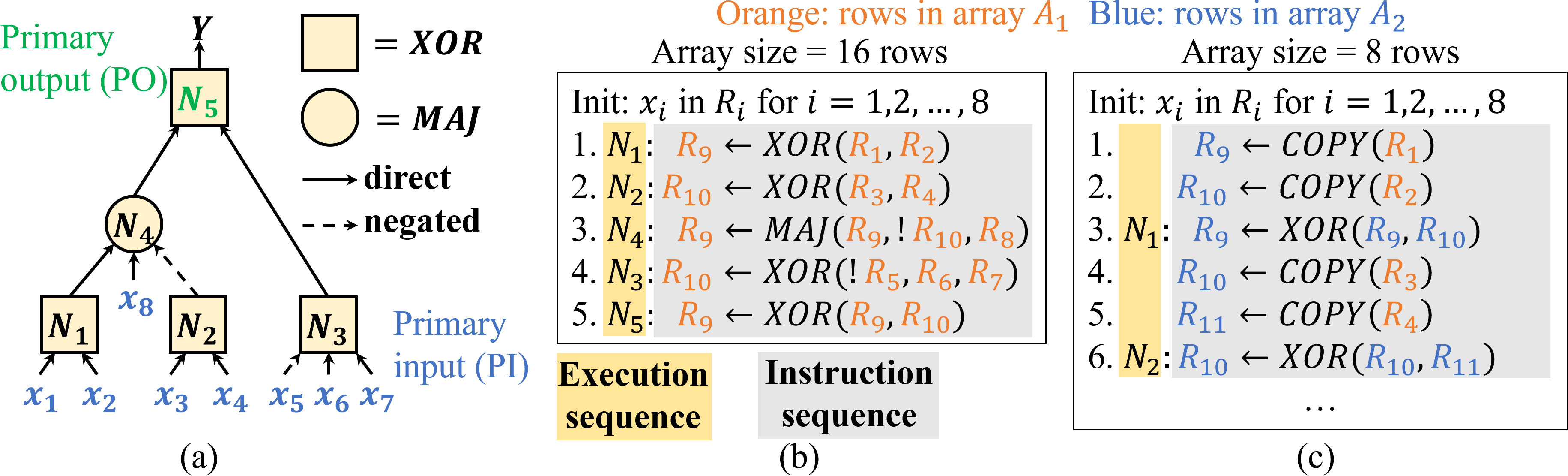}
    \caption{An example of (a)\qadd{ a} netlist\qdel{;}\qadd{,} (b)\qadd{ an} instruction sequence using a single array\qdel{;}\qadd{, and} (c)\qadd{ an} instruction sequence using \qdel{multiple-arrays}\qadd{multiple arrays}.}
    \label{fig:intr_sched}
\end{figure}

To implement a target function on the SIMD IMC architecture, the function is first transformed into a netlist by a logic synthesis tool. A netlist example is shown in Fig.~\ref{fig:intr_sched}(a), in which each node corresponds to a supported logic operation.
Then, a scheduler transforms the netlist into an \emph{instruction sequence (IS)} that instructs the architecture to implement the function\qadd{. An example of an IS is shown in the grey box}\qdel{ as shown} in Fig.~\ref{fig:intr_sched}\xdel{(b)}(c).
In SIMD setup, an instruction is given to the rows of the memory arrays, and it is either \qdel{computing a node, \textit{i.e.},}\qadd{performing a} logic operation\qdel{,} in the netlist\qdel{, \textit{e.g.}, using $R_9\leftarrow \textit{XOR}(R_9,R_{10})$ to compute $N_1$,} or copying a row of an array to a row of another array\qdel{, using $R_9\leftarrow \textit{COPY}(R_1)$ to copy fanin $x_1$ from $R_1$ in array $A_1$ to $R_9$ in array $A_2$}. 
For example, in Fig.~\ref{fig:intr_sched}(c), the instruction $R_9\leftarrow \textit{XOR}(R_9,R_{10})$ is an example of the former, which performs the XOR operation $N_1$, while the instruction $R_9\leftarrow \textit{COPY}(R_1)$ is an example of the latter, which copies fanin $x_1$ from $R_1$ in array $A_1$ to $R_9$ in array $A_2$.
\qadd{In what follows, we will also refer to performing a logic operation in the netlist as \emph{computing a node} in the netlist.}

Ideally, we want to use a single array to calculate the function as shown in Fig.~\ref{fig:intr_sched}(b)\xdel{.
In this case}, where each instruction in the IS \qdel{is to compute}\qadd{computes} a node in the netlist.
\xdel{Hence}In this case, most existing schedulers aim to reduce the number of rows needed during the calculation so that we can implement larger netlists within a single array with limited rows~\cite{OPTISIMPLER,MIG,XMG-GPPIC,SIMPLER,STAR,PREV}.
These schedulers focus on optimizing the \emph{execution sequence (ES) of the nodes}.
\qadd{An example of an ES of the nodes is the sequence of nodes $N_1,N_2,N_4,N_3,N_5$ shown in the yellow box of Fig.~\ref{fig:intr_sched}(b).}
\qadd{For simplicity, in what follows, we refer to an ES of the nodes as \emph{an ES}.}
We can see from Fig.~\ref{fig:intr_sched}(b) that to generate the \qdel{corresponding IS given an}\qadd{IS corresponding to a given} ES, we need to select a row for each node to store its result in.
We call the process that generates the corresponding IS given an ES the \emph{instruction generation (IG) process}.
The IG process \qdel{that targets at minimizing}that minimizes the number of rows needed \qdel{is easy}can by easily achieved by generating\qdel{.
It generates} one instruction for
each node to compute it and store its result in the available row with
the smallest index~\cite{PREV}.

\begin{wrapfigure}{r}{0.5\linewidth}
  \begin{center}
    \vspace{-1.5em}
    \includegraphics[width=\linewidth]{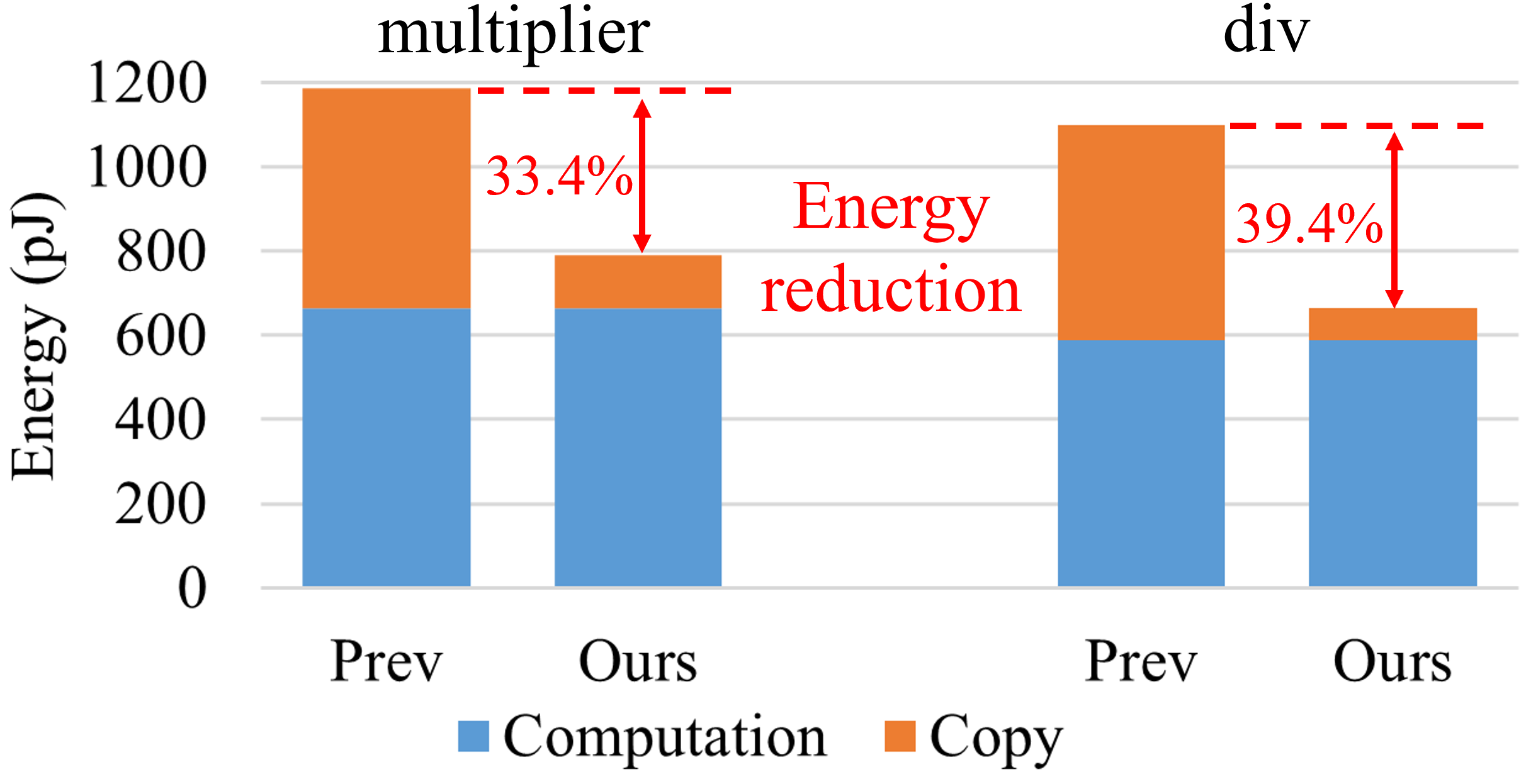}
  \end{center}
  \vspace{-1em}
  \caption{Two examples of energy \ndel{reduction}consumption of the same function implemented using\qadd{ the best prior} \qdel{previous works}\qadd{method} and our proposed method.}
  \vspace{-1.5em}
  \label{fig:mot}
\end{wrapfigure}

However, when the number of needed rows exceeds the size of the array, we have to use multiple arrays as shown in Fig.~\ref{fig:intr_sched}(c).
\qdel{Since}\qadd{For SIMD IMC,} a computation instruction can only be applied to the rows within the same array~\cite{XMG-GPPIC}.
Thus, we need to insert copy instructions to copy the data from one array to another.
However, a copy instruction is very energy-consuming: its energy consumption is 1.87 times that of a computation instruction\xdel{ as shown in Fig.~\ref{fig:mot}(a)}~\cite{XMG-GPPIC}.
Thus, the insertion of copy instructions can diminish the benefits offered by SIMD IMC~\cite{XMG-GPPIC,SIMPLER,CHAL}.

Hence, it is important to reduce the number of copy instructions.
However, existing schedulers do not take copy instructions into account when determining the \xdel{execution sequence}\xadd{ES}\xdel{, and row assignment and the insertion of copy instructions are determined in a naive~\cite{XMG-GPPIC} or greedy way~\cite{PREV}.}.
Given an ES, they use a naive~\cite{XMG-GPPIC} or greedy approach~\cite{PREV} to obtain the corresponding IS \xdel{in \qdel{the}\qadd{an} \emph{instruction generation (IG) process}}\nadd{in the IG process}\qdel{ that}\qadd{, which} \nadd{not only }selects the row to store the result of each node, \ndel{and}\nadd{but also} inserts necessary copy instruction \ndel{to compute}\nadd{before computing} the node.
In consequence, the resulting \xdel{instruction sequence}\xadd{IS} cannot minimize the number of copy instructions, and much energy is spent on copying data between the arrays.
\qdel{In}\qadd{As shown in Fig.~\ref{fig:mot}, in} our experiment on two frequently used functions, \qdel{\textit{i.e.}, }multiplication and division, the energy spent on copying is almost as large as that spent on computation using the best scheduling result \qdel{of}\qadd{from} the \xdel{best }previous works~\cite{XMG-GPPIC,SIMPLER,STAR,PREV}\qdel{ as shown in Fig.~\ref{fig:mot}\xadd{(a)}}.

\qadd{To address the above issue,}\qdel{In} this work\qdel{, we propose}\qadd{ proposes} \MethodName{}, a \underline{m}ulti-\underline{a}rray \underline{s}cheduler for \underline{i}n-\underline{m}emory SIMD computation, \xdel{a priority-driven\xdel{ list} scheduling algorithm }\qdel{that}\qadd{which} aims at reducing the number of copy instructions given an array size limit.
By reducing the number of copy instructions, the overall energy can be \qdel{greatly}\qadd{significantly} reduced as shown in Fig.~\ref{fig:mot}.
Our main contributions are as follows.
\begin{itemize}
\item 
We define a metric called the number of close partner pairs (\PName{s}), which can give a hint on the number of copy instructions needed in the future\qadd{ to help the scheduler make more judicious decision}.
We further propose a priority called \CombPName{}, which combines the number of \underline{c}opy \underline{i}nstructions and the number of \underline{\PName{}}s (see Section~\ref{sec:method_pridef}). \qadd{It serves as a key metric in \MethodName{}.}
\item 
\ndel{The algorithm}With \CombPName{}, we propose a \CombPName{}-driven scheduling algorithm, which can efficiently determine the \xdel{execution sequence}ES while taking reducing copy instruction number into consideration, and the \xdel{instruction sequence}IS is also obtained during the process (see Section~\ref{sec:method_sched}).
\item 
To further reduce \qadd{the }copy instruction number, we propose an iterative improvement \xdel{method.
The method}process, which randomly perturbs the \ndel{obtained }\xdel{execution sequence}ES and then \xdel{re-generates the instruction sequence}obtains the updated IS with a \CombPName{}-driven IG strategy, which also exploits the \CombPName{} priority (see Section~\ref{sec:method_iter})\ndel{ using the same\xadd{ \CombPName{}} priority}.
\item 
We also make the code of \MethodName{} open-source at 
https://github.com/SJTU-ECTL/MASIM.
\end{itemize}

The experimental results show that \xdel{our scheduler}\MethodName{} reduces the copy instruction number by 63.2\% on average compared to the best existing scheduler, which leads to a 28.0\% reduction in energy.

\section{Background}\label{sec:back}

\subsection{\qdel{Data Layout of SIMD IMC and Hardware \ndel{Limitation}Property}\qadd{SIMD IMC}}
\begin{figure}[!htbp]
    \centering
    \includegraphics[scale=0.32]{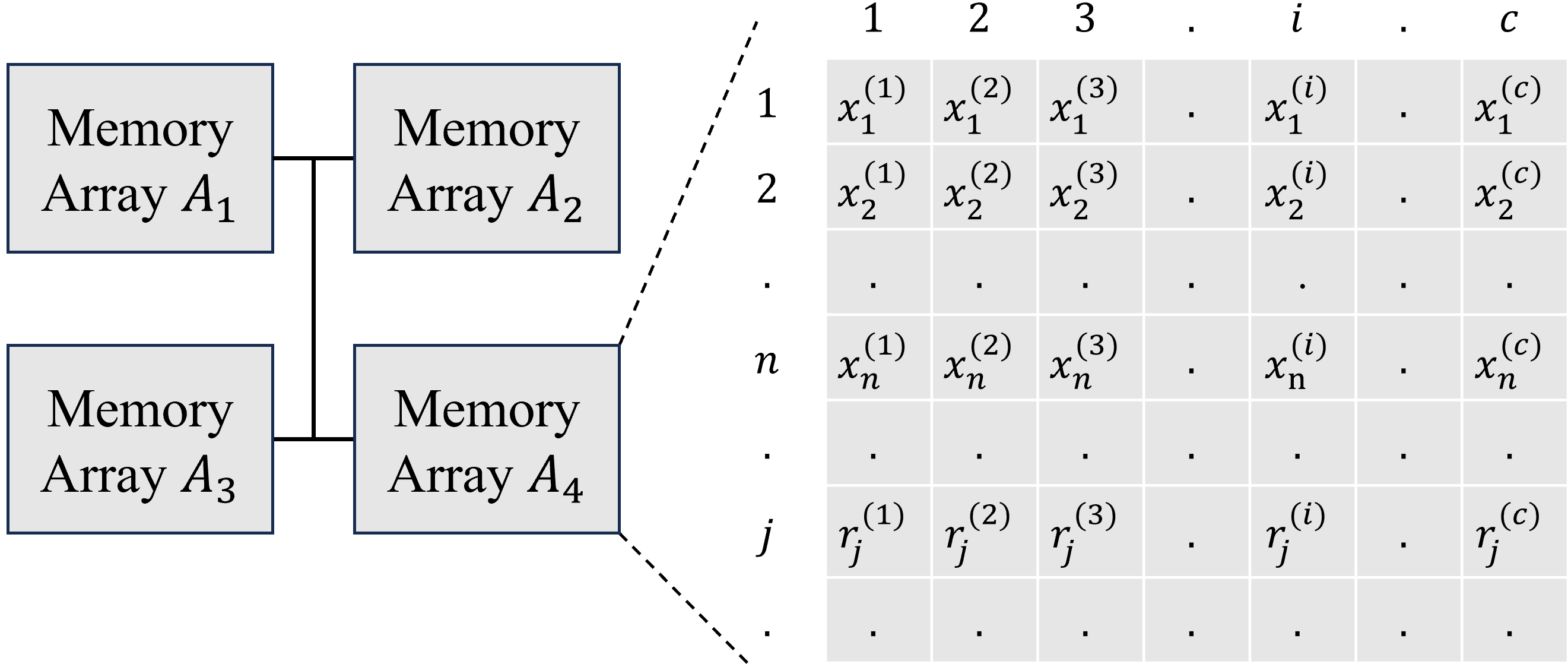}
    \caption{Data layout in memory array\qadd{s}.}
    \label{fig:data}
\end{figure}

SIMD is a popular IMC design style with high parallelism.
Fig.~\ref{fig:data} shows the data layout in a memory array with $r$ rows and $c$ columns.
In the SIMD setup, the array can calculate \xdel{$c$ functions}the same function $f$ on $c$\qadd{ different} input patterns\qadd{ $X^{(1)}, \ldots, X^{(\ndel{n}\nadd{c})}$} in parallel, where\qadd{ $X^{(i)} = (x_1^{(i)},...,x_n^{(i)})$ is the $i$-th $n$-bit input pattern, and the $i$-th output pattern $Y^{(i)}$, where $ Y^{(i)} = f(X^{(i)})$, is calculated in column $i$.}\qdel{ $Y^{(i)}=f(X^{(i)})$ is calculated in column $i$.}
Initially, the $n$-bit input pattern $X^{(i)} = (x_1^{(i)},...,x_n^{(i)})$ is placed in the topmost $n$ rows of column $i$\qdel{, and}\qadd{ as shown in Fig.~\ref{fig:data}. Then,} the function is calculated by passing a sequence of instructions to the hardware as shown in Fig\qadd{s}.~\ref{fig:intr_sched}(b) and\qdel{ Fig.}~\ref{fig:intr_sched}(c).
The instructions are given to the rows of the arrays, and the data stored in different columns of the same row are processed bitwise in parallel.
For example, by passing\qadd{ the instruction} $R_j\leftarrow \textit{XOR}(R_1,R_2)$ to the array in Fig.~\ref{fig:data}, $r_j^{(i)}=\textit{XOR}(x_1^{(i)},x_2^{(i)})$ is computed for each $i=1,...,c$.

However, an operation, \textit{e.g.}, $\textit{XOR}$ and $\textit{MAJ}$, can only be applied to the rows within the same array, whose size is \qdel{constrained by the leakage current paths originating from unselected cells, the decoder's driving ability, and other factors}\qadd{limited}~\cite{NVSIM}.
In the example in Fig.~\ref{fig:intr_sched}(c), \ndel{if}we assume that the array\qadd{s} only \qdel{contains}\qadd{have} 8 rows\ndel{, and the rows are labeled consecutively, \textit{i.e.}, $R_1$ to $R_8$ are in array $A_1$ and $R_9$ to $R_{16}$ are in array $A_2$, \textit{etc}}.
\ndel{\qadd{then }we}\nadd{We} cannot pass\qadd{ the instruction} $R_9\leftarrow \textit{XOR}(R_1,R_2)$ to \qadd{the }hardware directly since the rows $R_1$ and $R_2$ are \qdel{from}\qadd{in} array $A_1$, while \qadd{the row }$R_9$ is \qdel{from}\qadd{in} array $A_2$.
Instead, we need to insert copy instructions to copy the data in $R_1$ and $R_2$ to array $A_2$ before computing the operation in array $A_2$ \qdel{as shown}\qadd{(see instructions like $R_9 \leftarrow \textit{COPY}(R_1)$} in Fig.~\ref{fig:intr_sched}(c)\qadd{)}.
\ndel{In \qdel{the }architectures such as XMG-GPPIC~\cite{XMG-GPPIC}, the arrays are arranged hierarchically.
Typically, 4 arrays form a \emph{mat} and four mats form a \emph{bank}.
Cross-mat copy instructions are much more \qdel{energy consuming}\qadd{energy-consuming} than cross-array copy instructions within a mat, so we focus on scheduling within a mat and consider the energy consumption of a copy instruction to be constant.}

\subsection{Netlist Representation and Scheduling Rules of SIMD IMC}\label{sec:back_rule}
As mentioned before, to fully exploit the potential of the SIMD IMC architecture, a scheduler is needed to automatically transform the netlist of the target function synthesized by the logic synthesis tool to the \xdel{instruction sequence}\xadd{IS} given to the hardware.
\qdel{In the illustration and the experiment}\qadd{In this work,} we use\xdel{Our target hardware is} XMG-GPPIC proposed in~\cite{XMG-GPPIC}, which supports 3-input $\textit{XOR}$ and $\textit{MAJ}$ operation with fused negation, as the target hardware.
Therefore, the corresponding netlist is an XOR-majority graph (XMG).
\qdel{Note that}\qadd{However, note that} our method can be applied to other hardware with some minor modifications.
\qdel{We will}\qadd{Next, we} explain the component of the XMG together with the scheduling rules using the \ndel{example }XMG\qdel{ as} shown in Fig.~\ref{fig:intr_sched}(a)\qdel{:}\qadd{.}
\begin{itemize}
\item
The primary inputs (PIs), which are shown in blue in Fig.~\ref{fig:intr_sched}(a), are initially stored in the topmost rows of the array.
That is, the input $x_i$ is stored in \qdel{the }row $R_i$ for $i=1,2,...,8$.
Since the PIs may be needed in the future when calculating other functions, they cannot be overwritten during the process.
\item
Each node in the netlist corresponds to a supported logic operation, \textit{i.e.}, either\qadd{ a} 3-input $\textit{XOR}$ or\qadd{ a} 3-input $\textit{MAJ}$.
Note that in the figure, some operations only have \qdel{2-}\qadd{2 }inputs\qadd{,} since \qadd{the missing input is a }constant \qdel{inputs}\qadd{$0$ or $1$, which} \qdel{are}\qadd{is} omitted for simplicity.
In the SIMD setup, the operations are computed sequentially, and it takes one clock cycle to compute \ndel{each}\nadd{an} operation.
\item
The directed edges indicate the dependency relation.
If there is a directed edge from node $i$ to node $j$, \qadd{then }the output of node $i$ is the input of node $j$.
If the edge is dashed, the output of node $i$ is negated when it serves as the input of node $j$.
We call node $i$ a \emph{fanin} of node $j$ and node $j$ a \emph{fanout} of node $i$.
A node can be computed in an array if and only if all its fanins are stored in that array\qdel{, and}\qadd{. Furthermore,} the hardware allows the output of a node to overwrite one of its inputs.
\item
The results of the primary outputs (POs), which are shown in green in Fig.~\ref{fig:intr_sched}(a), must be stored in the array when the schedule ends\qdel{, and}\qadd{. However,} the results of\qadd{ the} other nodes can be overwritten when not needed any more.
\end{itemize}

\subsection{Existing Schedulers}\label{sec:back_exist}
Many schedulers have been proposed for SIMD IMC architectures.
Most of them focus on the computation within a single array.
They aim to reduce the number of rows needed during the \qdel{calculation}\qadd{computation} so that we can implement larger netlists within a single array with a limited size~\cite{OPTISIMPLER,MIG,SIMPLER,STAR,PREV}.
In order to achieve this, \qdel{the}\qadd{their} main effort is\nadd{ to} \qdel{made in determining}\qadd{determine} the \xdel{execution sequence}\xadd{ES}\ndel{ of the nodes}.
Given an \xdel{execution sequence}\xadd{ES},\qadd{ in order to use fewest rows,} the row \xdel{assignment}to store the\qadd{ computation} result of each node\xdel{ that needs the fewest rows} can be \qdel{easily}\qadd{simply} \xdel{determined}selected \ndel{by overwriting the results of the nodes that are no longer needed}\qdel{to be}\qadd{as} the available row with the smallest index\qdel{ to achieve the minimum number of rows needed}~\cite{PREV}.
The \xdel{instruction sequence}corresponding IS can be easily determined by computing one node \qdel{at each}\qadd{per} clock cycle according to the \xdel{execution sequence}\xadd{ES} and storing its result in the \xdel{assigned}\qdel{selected}\qadd{chosen} row without the need of copy instructions.

To determine the ES with \qdel{minimum number of}\qadd{the fewest} rows needed, many methods have been proposed.
OptiSIMPLER can obtain the \xdel{execution sequence}\xadd{ES} with the optimal row requirement for the given netlist by formulating a Boolean satisfiability (SAT) problem~\cite{OPTISIMPLER}. However, since SAT solving is time-consuming, the scheduler can only be applied to small netlists.
To support large netlists, other schedulers turn to heuristic approaches.
Some schedulers are based on priority-driven \xdel{list }scheduling.
At each clock cycle, a node that is ready to be computed is picked according to a priority.
The priority in the scheduler in~\cite{XMG-GPPIC} is adapted from that in~\cite{MIG}\xdel{.
A node with more releasing fanins and less level of fanouts is favored to minimize the row requirement.}, where a node that can remove more fanins from memory after scheduling it is favored.
In STAR~\cite{STAR}, \xdel{the minimal number of nodes to be computed before erasing a node from the memory is used as the priority}the priority\qadd{ of a node $n$} is determined by the minimum number of nodes that need to be computed before \qdel{a node}\qadd{node $n$} can be removed from memory.
SIMPLER~\cite{SIMPLER} performs a depth-first search on the netlist from the outputs until reaching a node that is ready to \ndel{compute}be computed.
The traversal order \ndel{priority }\qdel{for}\qadd{of the} fanins is determined by an optimistic estimation of the number of rows needed to calculate the fanins, \textit{i.e.}, Strahler number~\cite{STRAHLER}.
Another recent work uses a divide-and-conquer approach that partitions a large netlist into small sub-netlists that can be scheduled by an optimal scheduler~\cite{PREV}.
The scheduler outperforms previous ones in reducing the \xdel{row requirement}number of rows needed at the cost of a longer runtime since the solvers~\cite{Z3,GUROBI} used in the scheduler are time-consuming.

Some existing schedulers also report their experimental results in terms of energy and delay when using multiple arrays~\cite{XMG-GPPIC,PREV}.
However, they do not make much effort to reduce the number of copy instructions, so the resulting \xdel{instruction sequence}\xadd{IS} spends \qdel{a lot of}\qadd{much} energy \qdel{on}\qadd{in} copying as shown in Fig.~\ref{fig:mot}.
Based on the obtained ES, XMG-GPPIC uses a naive \xdel{copy instruction generation}\xadd{IG} strategy\ndel{ based on the obtained \xdel{execution sequence}\xadd{ES}} to select the row to store the result of each node and inserts necessary copy instruction\ndel{ to compute the node}~\cite{XMG-GPPIC}.
Specifically, the array with the smallest index that contains enough available rows is selected to compute each node according to the ES.
Before\qadd{ issuing} the computation instruction, copy instructions are inserted to copy each fanin\qdel{, which}\qadd{ that} is not stored in the selected array to \qdel{one}\qadd{an} available row in the array.
The recent work improves the naive strategy to a greedy one~\cite{PREV}.
\xdel{For each node, the row assignment is chosen}The array used to compute a node is selected so that the number of copy instructions inserted to copy its fanins is minimized.
However, since the \xdel{execution sequence}\xadd{ES} is determined to minimize the \xdel{row requirement}number of rows needed instead of the number of copy instructions, the result can still be \qdel{greatly}\qadd{much} improved.

\section{Method}\label{sec:method}
\subsection{Overview}\label{sec:method_over}
As mentioned \qdel{before}\qadd{in Section~\ref{sec:back_rule}}, a scheduler \qdel{automatically }transforms the netlist of the target function \ndel{to}\nadd{into} the \xdel{instruction sequence}IS given to the hardware.
The instructions are executed by the hardware in serial with one instruction per clock cycle.
Hence, the \qdel{overall}\qadd{total} energy for implementing\qdel{computing}\qdel{calculating} the target function \ndel{on the hardware }is \qdel{the result of }the accumulation of the energy used to execute each instruction.
\qdel{Since}\qadd{To minimize the energy,} we\qadd{ typically} only allow computing each node \qadd{(\textit{i.e.}, performing the corresponding logic operation) }in the netlist \emph{once}\qdel{ to minimize the energy}\qadd{. Thus}, the total energy spent on executing all the computation instructions is fixed for a given netlist.
Therefore, to minimize the \qdel{overall}\qadd{total} energy consumption, it is important to \emph{minimize the total energy spent on executing all \qdel{the }copy instructions}\qadd{. In this work, we assume that the energy consumption of a copy instruction is constant.
Thus, our target reduces to \emph{minimize the number of copy instructions for a given netlist and a given array size}.}\qdel{, which is proportional to the number of copy instructions, assuming \qadd{that }the energy consumption of a copy instruction to be\qadd{ a} constant.}

\begin{figure}[!htbp]
    \centering
    \includegraphics[scale=0.35]{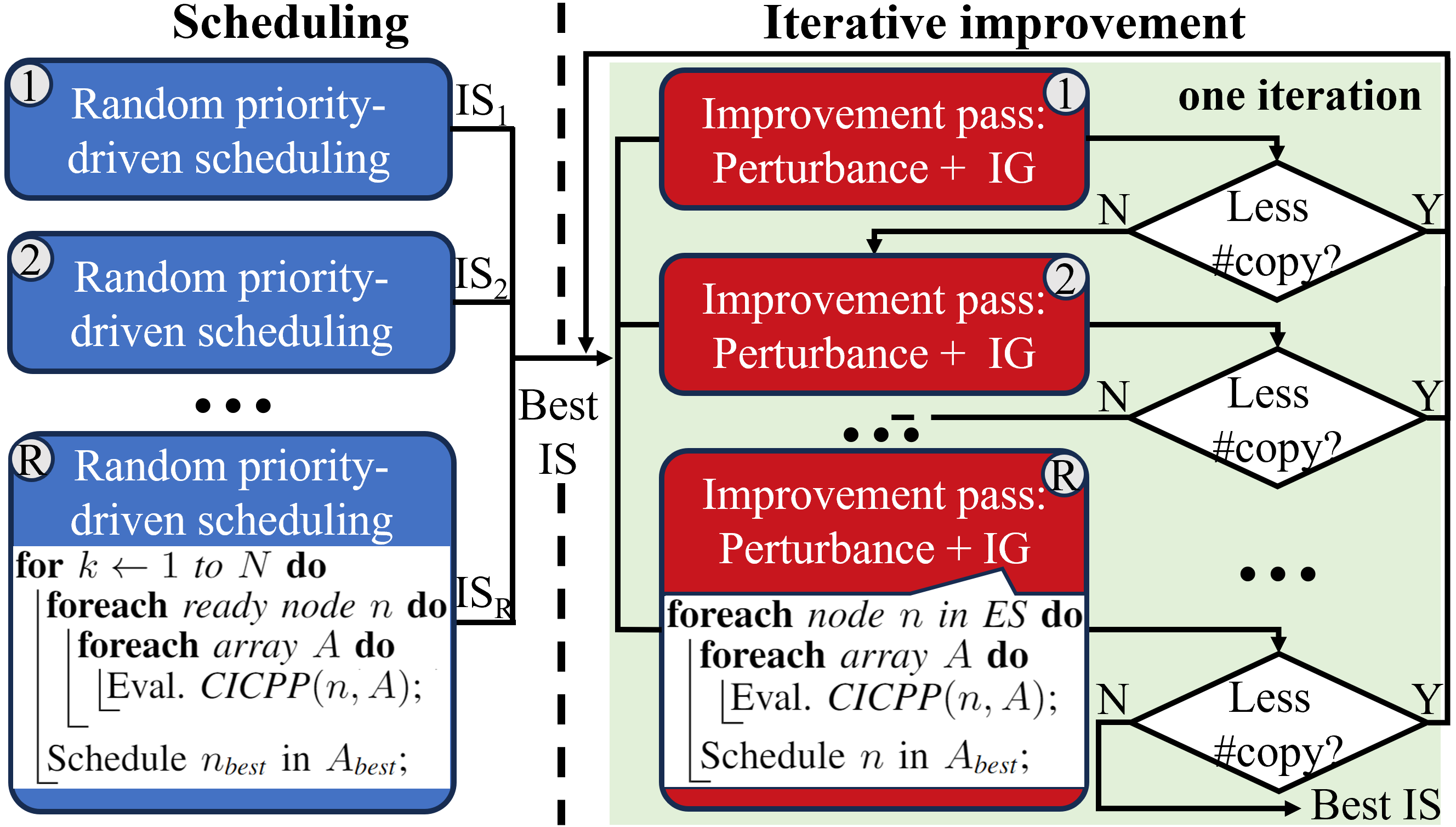}
    \caption{\qdel{Flow chart of overview process}\qadd{The overall flow of \MethodName{}}.}
    \label{fig:overv}
\end{figure}

\qdel{We take minimizing copy instruction number}\qadd{Our proposed solution, \MethodName{}, takes minimizing the copy instruction number} into account when determining the \qdel{\xdel{execution sequence}\xadd{ES} of the nodes, which is different from the existing methods}\qadd{IS}.
\qdel{The overview of \xdel{our method}\MethodName{}}\qadd{Its overall flow} is shown in Fig.~\ref{fig:overv}, which consists of two main parts, \textit{i.e.}, a random \xdel{scheduler}priority-driven scheduling algorithm and an iterative improvement \xdel{technique}process.

In the first part, our scheduling algorithm needs to make the following three decisions iteratively until all nodes are scheduled: 1)~\qadd{\textbf{node choice}, \textit{i.e.},} which node\qdel{ $n$} to be scheduled next\qdel{, determining the ES}, 2)~\qadd{\textbf{array choice}, \textit{i.e.},} which array\qdel{ $A$} to schedule the node in, and 3)~\qadd{\textbf{row choice}, \textit{i.e.},} which rows in\qadd{ the} array\qdel{ $A$} are used to store the \qdel{\emph{incoming data}, \textit{i.e.}, }missing fanins and the result of the node.
We call the missing fanins and the result of the node \emph{incoming data}.
\ndel{Note that we}We say a fanin $i$ of $n$ is \emph{missing} in array $A$ if $i$ is not stored in array $A$.
In \MethodName{}, we use \qdel{what we call}a metric called \CombPName{}, which will be introduced in Section~\ref{sec:method_pridef}, as the priority when making these decisions.
Once we \qadd{have }made the decisions, we generate \qadd{the }instructions to copy the missing fanins and then compute the \qdel{selected}\qadd{chosen} node accordingly.
When there is a tie in priority, the scheduler picks a random choice.
\qadd{Hence, we call the scheduler \emph{random \CombPName{}-driven scheduler}.}
\qdel{The random \CombPName{}-driven scheduler}\qadd{As shown in Fig.~\ref{fig:overv}, the random scheduler} is run for $R$ times with different random seeds\qadd{ used in breaking the ties} to obtain the best \xdel{instruction sequence}\xadd{IS}, \textit{i.e.}, the one with the minimum number of copy instructions.
The details of the \ndel{scheduler}scheduling algorithm will be introduced in Section~\ref{sec:method_sched}.

In the second part, we attempt to iteratively improve the result.
\qdel{In each}\qadd{One} iteration\qadd{ of this iterative process is} shown in \qadd{the }green box in Fig.~\ref{fig:overv}\qdel{,}\qadd{. In each iteration,} we attempt to decrease the copy instruction number with at most $R$ \qdel{\qdel{runs}\qadd{trials} of \emph{improvement process}}\qadd{\emph{improvement passes}}\qdel{ as shown in the}\qadd{, each shown as a} red box\ndel{ in Fig.~\ref{fig:overv}}.
In each \qdel{run of the }improvement \qdel{process}\qadd{pass}, we first randomly \qdel{disturb}\qadd{perturb} the \xdel{execution sequence of the nodes}\xadd{ES} in the best \xdel{instruction sequence}\xadd{IS}\qadd{ obtained} so far.
Then, we apply an \xdel{instruction generation (IG)}\xadd{IG} process based on the new \xdel{execution sequence}\xadd{ES} using the same \CombPName{} priority.
Note that the IG process only needs to make the last two decisions of the scheduling decisions\qadd{, \textit{i.e.}, array choice and row choice,} since the \qdel{ES of the nodes, which is decided in decision 1), is given}\qadd{node choice is decided by the given \xadd{ES}}.
If the resulting IS has fewer copy instructions, we accept it\qadd{ as \emph{the best IS obtained so far}} and begin a new iteration of the iterative improvement process.
Otherwise, we begin another \qdel{\qdel{run}\qadd{trial} of the improvement process}\qadd{improvement pass} in the current iteration.
If none of the results in \qdel{the }$R$ \qdel{runs}\qadd{\ndel{consecutive }improvement passes} in an iteration is accepted, we end the flow and output the updated IS.
The details of the \ndel{iterative improvement process}improvement pass will be \qdel{introduced}\qadd{described} in Section~\ref{sec:method_iter}.
\subsection{Random \CombPName{}-Driven Scheduling Algorithm}\label{sec:method_sched}
\begin{figure}[!htbp]
    \centering
    \includegraphics[scale=0.32]{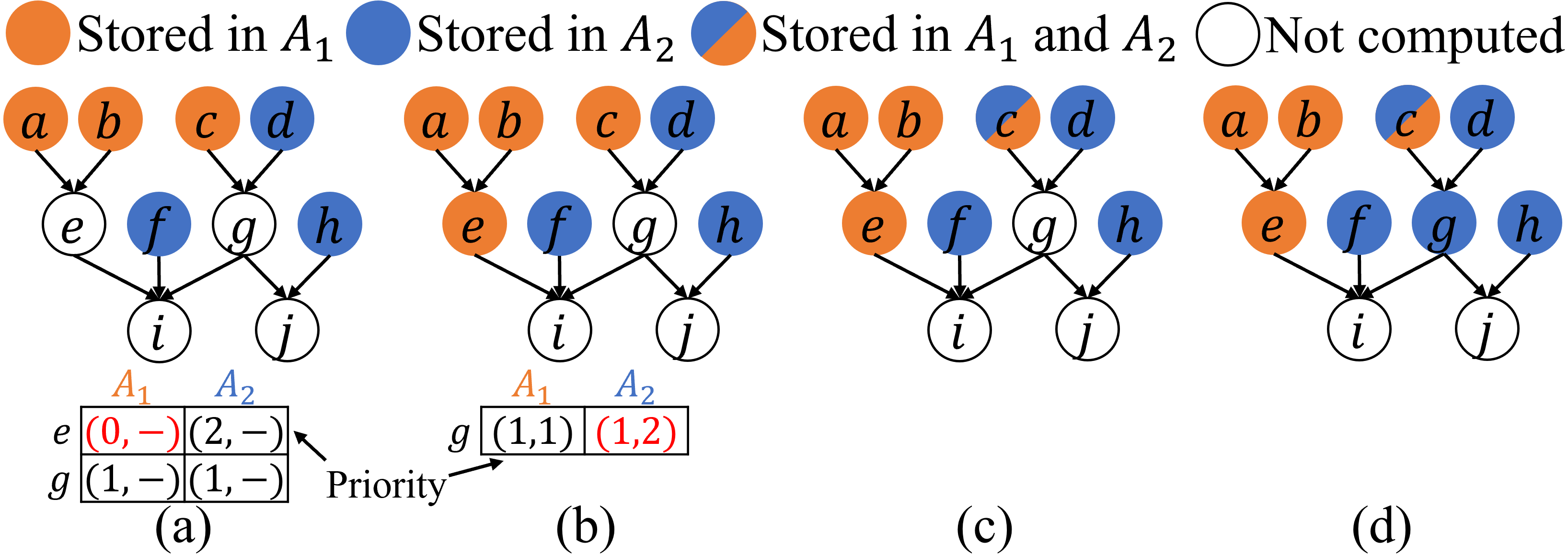}
    \caption{Scheduling example: (a) \qadd{the }initial state\qadd{ of an outermost iteration of the random priority-driven scheduling}; (b) compute $e$ in $A_1$; (c) copy $c$ to $A_2$; (d) compute $g$ in $A_2$.}
    \label{fig:sched}
\end{figure}
The foundation of our scheduling algorithm is a priority-driven scheduler.
We say that a node is \emph{ready} if and only if all its fanins have already been computed and thus stored in the arrays.
The overview of our scheduling algorithm is shown in \ndel{left-bottom corner of}the \qdel{last }blue box \qadd{at the bottom-left corner of}\qdel{in} Fig.~\ref{fig:overv}.
\qadd{The outermost loop has $N$ iterations}\qdel{We iteratively schedule\qadd{s} the nodes in netlist until all $N$ nodes have been scheduled}, where $N$ is the number of nodes\qadd{ in the given netlist}.
\qadd{Each iteration of it decides the node choice, the array choice, and the row choice.}
\qadd{Thus, after the loop finishes, all $N$ nodes have been scheduled.}
\qadd{To make the above three decisions in each iteration, we traverse}\qdel{For} each ready node $n$ and\qdel{ for} each array $A$\ndel{,} \qdel{we}\ndel{and call the function $\textit{CICPP}$ on node $n$ and arary $A$, which evaluates}and evaluate the \CombPName{} priority of scheduling $n$ in $A$, \qdel{\textit{i.e.},}\qadd{denoted as} $\textit{CICPP}(n,A)$\ndel{ and returns the priority together with the row choice obtained in the evaluation process}.
Note that the corresponding row choice can be also obtained in the evaluation process to be illustrated in Section~\ref{sec:method_prieval}.
Based on the evaluation results\qadd{,} we choose to schedule node $n_{\textit{best}}$ in array $A_{\textit{best}}$ with the highest $\textit{CICPP}(n_{\textit{best}},A_{\textit{best}})$ according to the corresponding row choice.
That is, we copy the missing fanins of $n_\textit{best}$ to the selected rows and then compute $n_\textit{best}$, putting its result in another selected row.
\vspace{-1em}
\begin{Example}
\qdel{In the}\qadd{Consider an} example in Fig.~\ref{fig:sched}(a)\qadd{. We want to decide the node and array choices for the next iteration.}
\qdel{,}\qadd{In the current state shown in Fig.~\ref{fig:sched}(a),} node\qadd{s} $e$ and \qdel{node }$g$ are ready since \qdel{their fanins, \textit{i.e.}, nodes}\qadd{the fanins of $e$, \textit{i.e.}, nodes $a$ and $b$, and the fanins of $g$, \textit{i.e.}, nodes $c$ and $d$,} \qdel{$a,b,c,d$ }have been computed and are stored in the arrays\qdel{, and}\qadd{. We assume that the hardware has two arrays, $A_1$ and $A_2$.}
\qdel{we}\qadd{We} evaluate the priority of scheduling each \qdel{of them}\qadd{ready node (\textit{i.e.}, $e$ and $g$)} in\qadd{ each} array \qdel{$A_1$ or $A_2$}\qadd{(\textit{i.e.}, $A_1$ and $A_2$)}.
The evaluated priority is shown in the table in the figure.
Among four possible choices, we choose to schedule node $e$ in array $A_1$, which gives the highest priority marked in red.
\qadd{The definition of the priority and the detailed values of the priorities in the table will be shown later.}
\end{Example}

\qdel{For now}\qadd{In the following}, we\qadd{ will first} assume that the arrays have enough \ndel{free}\emph{empty rows}\qadd{ that contain no data} so that we do not need to make the row choice since we can use arbitrary rows to store the incoming data.
Based on this assumption, we\qadd{ will} explain the basic idea of \CombPName{} priority in Section~\ref{sec:method_pridef}.
\qadd{Then, in}\qdel{In} Section~\ref{sec:method_prieval}, we will illustrate how to evaluate \CombPName{} priority when taking\qadd{ the} row choice into account\ndel{ and the row selection strategy used in the evaluation is explained in Section~\ref{sec:method_row}}.

\subsubsection{Basic Idea of \CombPName{} Priority}\label{sec:method_pridef}
In this section, we introduce the basic idea of our \CombPName{} priority under the assumption that the arrays have enough empty rows.
We set minimizing the number of copy instructions\qdel{ that are} needed to schedule the current node as our primary priority since it is the target of our scheduler.
\vspace{-1em}
\begin{Example}
In the example in Fig.~\ref{fig:sched}(a), among the four possible choices, scheduling node $e$ in array $A_1$ does not need any copy instruction, while the other choices need 1 or 2 copy instructions.
Thus, we decide to schedule $e$ in array $A_1$ by generating a computation instruction, and the status of the nodes \qdel{becomes}\qadd{after this step is shown in} Fig.~\ref{fig:sched}(b).
Note that in this case, \qdel{we can make the scheduling choices with the primary priority only}\qadd{the primary priority is enough to make the scheduling decision}, so the secondary priority to be introduced below does not matter, and \qdel{we put}\qadd{hence,} it is denoted as a ``--'' in the table in Fig.~\ref{fig:sched}(a).
\end{Example}

When there is a tie in the primary priority, we turn to our secondary priority\qdel{, \textit{i.e.}, maximizing the number of \todo{\emph{close partner pairs (\PName{s})}}\qdel{\PName{s}}, which can give a hint on the number of copy instructions needed in the future}.
\qadd{Before introducing it, we fist give a few definitions.}
We say that node $n$ and node $i$ are \emph{partners} of each other if they share a fanout $x$ that has not been computed yet.
If two partner nodes are stored in the same array $A$, we say that they form a \emph{close partner pair} (\PName{}) in array $A$.
In the example in Fig.~\ref{fig:sched}(a), node $a$ and node $b$ form a \PName{} in array $A_1$.
Although node $c$ and node $d$ are partners of each other, they do not form a \PName{} since they are stored in different arrays.
We propose to maximize the change in the number of \PName{s}, denoted as $\Delta\#\textit{CPP}$, as our secondary priority.
By maximizing $\Delta\#\textit{CPP}$ in each step, we can reach a memory status with more CPPs, which is favorable since the fanins of a node to be scheduled are more likely to be stored in the same array, thus requiring fewer copy instructions in the future, as shown in the following example.
\vspace{-1em}
\begin{Example}
In the example in Fig.~\ref{fig:sched}(b), only node $g$ is ready, and scheduling it \qadd{either }in array $A_1$ \qdel{and}\qadd{or in} array $A_2$ \qdel{both need}\qadd{needs} one copy instruction, so we need to use the secondary priority.
\nadd{If we schedule $g$ in $A_1$, $\Delta\#\textit{CPP}=1$ since node $g$ and node $e$ form a CPP in $A_1$.
If we schedule $g$ in $A_2$, $\Delta\#\textit{CPP}=2$ since node $g$ forms one CPP with node $f$ and\qadd{ one with} node $h$\qdel{, respectively,} in $A_2$.}
Therefore, we choose to schedule $g$ in $A_2$ since it has a larger \ndel{change in the number of \PName{s}}\nadd{$\Delta\#\textit{CPP}$} and potentially needs fewer copy instructions in the future.
Given this choice, we first copy the missing fanin $c$ into array $A_2$\qdel{, \textit{e.g.},}\qadd{ (see }Fig.~\ref{fig:sched}(c)\qdel{,}\qadd{)} and then compute $g$ in $A_2$\qdel{, \textit{e.g.},}\qadd{ (see} Fig.~\ref{fig:sched}(d)\qadd{)}.
Starting from the memory status in Fig.~\ref{fig:sched}(d), we need one copy instruction to compute node $i$ and node $j$, \textit{i.e.}, copying node $e$ to array $A_2$.
However, if we schedule $g$ in $A_1$, we will need two copy instructions.
This observation confirms the rationality of our secondary priority.
\end{Example}

\subsubsection{Evaluation of \CombPName{} Priority}\label{sec:method_prieval}
In this section, we illustrate the way to evaluate the \CombPName{} priority to schedule node $n$ in array $A$ when taking the row choice into account.
Since the rows we select will be used to store the incoming data, the original nodes stored in any of these rows will be overwritten, possibly leading to a reduction in the \PName{} number\qadd{ and hence, a negative $\Delta\#\textit{CPP}$}.

\qdel{To make it convenient to evaluate $\Delta\#\textit{CPP}$}\qadd{To facilitate our discussion}, we use $N_\textit{PA}(x,A)$ to denote the number of partners of node $x$ stored in array $A$.
By definition, when node $x$ emerges in array $A$ by copying or computing, the number of \PName{s} increases by $N_\textit{PA}(x,A)$.
When node $x$ is overwritten in array $A$, the number of \PName{s} decreases by $N_\textit{PA}(x,A)$.

As mentioned in Section~\ref{sec:method_over}, to schedule node $n$ in array $A$, we should first copy all the missing fanins to $A$.
For each missing fanin $i$, we select a row in $A$ and copy $i$ to that row.
\qadd{Note that it is possible that the selected row cannot be directly overwritten, which requires some additional copy instructions that may change the number of \PName{s}, which we will elaborate in Section~\ref{sec:method_row}.}
\qadd{Thus, the selection of a row to store the missing fanin $i$ will introduce some copy instructions and a change in \PName{} number.}
Suppose that the selection process needs $\#\textit{CI}_i$ copy instructions and the change in \PName{} number is $\Delta\#\textit{CPP}_i$.
After the row is selected, we copy $i$ to the row, which needs 1 copy instruction and increases the CPP number by $N_\textit{PA}(i,A)$.
Then, with all fanins stored in $A$, we compute $n$ in $A$.
We select a row in $A$ to store the result of $n$.
Suppose that the selection process needs $\#\textit{CI}_n$ copy instructions\qdel{,} and the change in \PName{} number is $\Delta\#\textit{CPP}_n$.
When computing $n$ and storing its result in the selected row, we need 0 copy instruction, and the CPP number increases by $N_\textit{PA}(n,A)$.
Throughout this process, the two priorities are accumulated.
That is, the total number of copy instruction needed\qadd{, which is the primary priority for scheduling node $n$ in array $A$,} is $\sum_i(\#\textit{CI}_i+1)+\#\textit{CI}_n$, and the total change in \PName{} number\qadd{, which is the secondary priority for scheduling node $n$ in array $A$,} is $\sum_i(\Delta\#\textit{CPP}_i+N_\textit{PA}(i,A))+\Delta\#\textit{CPP}_n+N_\textit{PA}(n,A)$.

\subsubsection{Row Selection Strategy}\label{sec:method_row}
\begin{figure}[!htbp]
    \centering
    \includegraphics[scale=0.3]{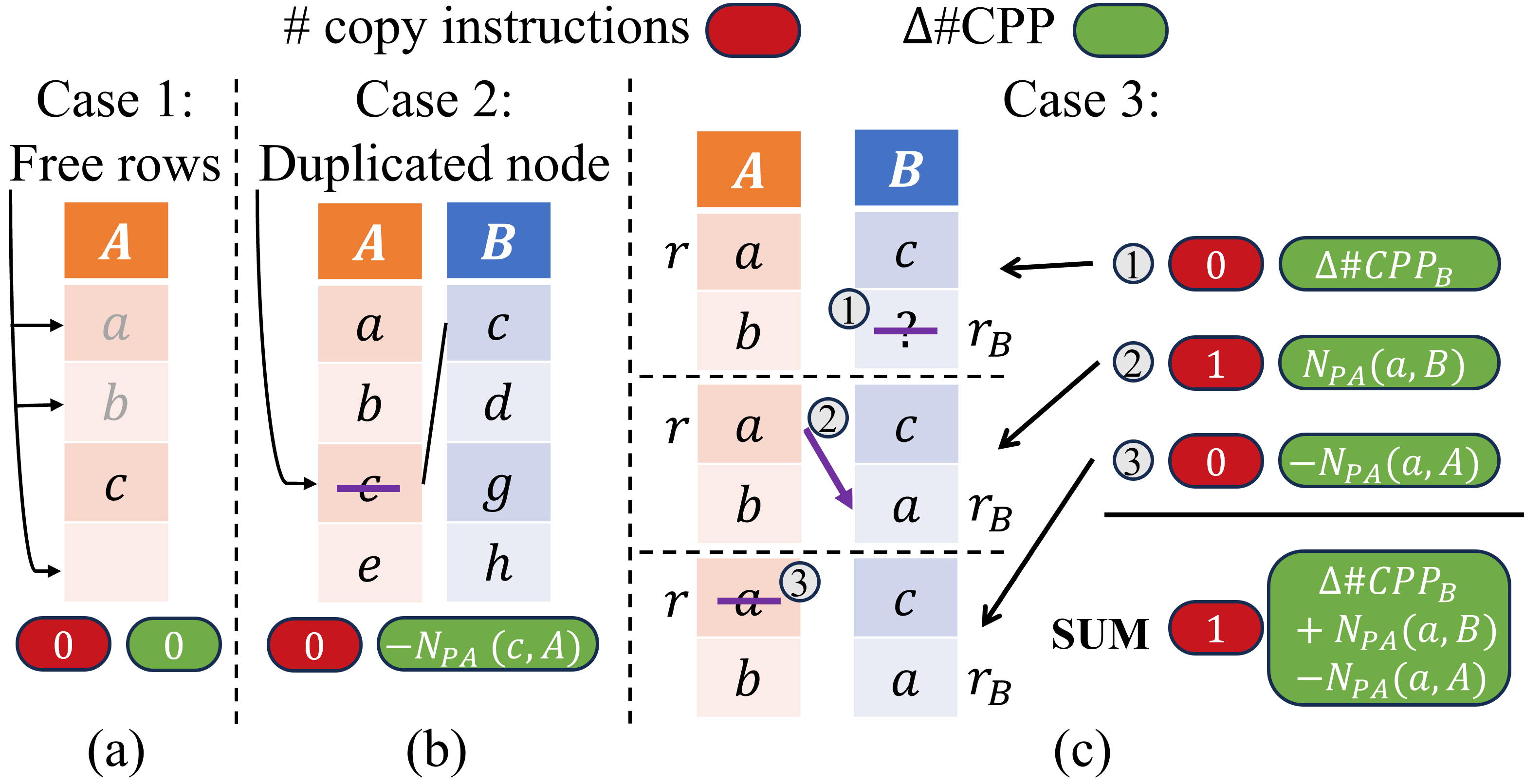}
    \caption{Row selection strategy: (a) Case 1; (b) Case 2; (c) Case 3.}
    \label{fig:row_sel}
\end{figure}
Now we illustrate our row selection strategy using Fig.~\ref{fig:row_sel}.
In the figure, the number of copy instructions is shown in the red circles and $\Delta\#\textit{CPP}$ is shown in the green circles.
As mentioned in Section~\ref{sec:method_prieval}, the number of copy instructions and $\Delta\#\textit{CPP}$ for a row selection will be accumulated in the total \CombPName{} priority, so the row selection is also done according to \CombPName{} priority, \textit{i.e.}, we attempt to minimize the number of copy instructions and maximize $\Delta\#\textit{CPP}$.
When we need to select a row in array $A$, the array belongs to one of the following three cases with decreasing priority.
We illustrate our strategy for each case.

\textbf{Case 1} is the ideal case where there are free rows in the array, as shown in Fig.~\ref{fig:row_sel}(a).
A row is said to be \emph{free} if it is empty or the stored node is no longer needed.
For example, in Fig.~\ref{fig:sched}(b), after computing node $e$ in array $A_1$, the results of node $a$ and node $b$ are no longer needed and can be overwritten without any potential disadvantage, so\nadd{ we mark them in grey in Fig.~\ref{fig:row_sel}(a), and} the rows storing their results are regarded as free rows.
In this case, \nadd{we select an arbitrary free row,  and }no copy instruction is needed and $\Delta\#\textit{CPP}=0$.

In \textbf{Case 2}, there is no free row in the array, but there are duplicated nodes stored in the array as shown in Fig.~\ref{fig:row_sel}(b).
A node is said to be \emph{duplicated} if 1) it is needed for the future computation, \textit{i.e.}, it is a PO or has fanouts that have not been computed yet, and 2) it is stored in more than one array.
For example, in Fig.~\ref{fig:sched}(c), node $c$ is duplicated since it has a fanout $g$ that has not been computed yet, and it is stored in both arrays $A_1$ and $A_2$.
Ideally, we avoid overwriting a duplicated node in an array, since we may need to compute its fanouts in that array.
For example, if we overwrite node $c$ in array $A_2$ with the computation result of \qdel{some }\qadd{an}other node\qdel{ $x$} before scheduling node $g$\nadd{ in Fig.~\ref{fig:sched}(c)}, $c$ will be missing when scheduling $g$, so extra copy instructions will be needed to compute $g$.
However, when there is no free row in an array, we have to select a row that stores a duplicated node and overwrite it\nadd{, which is indicated by the purple line in Fig.~\ref{fig:row_sel}(b)}.
When overwriting duplicated node $c$ in array $A$, the number of CPPs is reduced by $N_\textit{PA}(c,A)$, and among all duplicated nodes stored in array $A$, we choose to overwrite the one with\qadd{ the} minimum CPP number reduction.
In this case, still no copy instruction is needed, but $\Delta\#\textit{CPP}$ \qdel{might}\qadd{may} be a negative number\qadd{, which is less favorable}.

\textbf{Case 3} is the worst case where there is no free row and all nodes stored in the array\nadd{ $A$} are not duplicated, as shown in Fig.~\ref{fig:row_sel}(c).
Since the nodes are not duplicated, but will still be needed in the future, we cannot simply overwrite them.
Instead, if we want to select a row $r$ in array $A$, we first need to insert a copy instruction to copy the \qdel{stored }node $a$ \qadd{stored }in $r$ to another array $B$.
This \qdel{arise}\qadd{raises} \qdel{another}\qadd{a new} question: which row in $B$ should we use to store the copied $a$.
\qdel{Hence, selecting a row in $A$ in case 3 is}\qadd{We propose} a three-step \qdel{process}\qadd{procedure to select a row \nadd{$r$ }in $A$}.
In Step~\ding{172},\nadd{ as shown in Fig.~\ref{fig:row_sel}(c),} we select a row $r_B$ in another array $B\neq A$ using the \ndel{same }row selection strategy\nadd{ for Case 1 or 2}.
\nadd{That is, we select a free row or a row containing a duplicated node in $B$.}
If no row can be selected in this way, we move on to the next \qdel{$B$}\qadd{available array}.
In Step~\ding{172}, no copy instruction is needed and the change in CPP number is denoted as $\Delta\#\textit{CPP}_B$.
In Step~\ding{173}, we generate a copy instruction to copy $a$ \nadd{from row $r$ }to\nadd{ row} $r_B$\nadd{ as indicated by the purple arrow in Fig.~\ref{fig:row_sel}(c)}.
In this step, 1 copy instruction is needed and the change in CPP number is $N_\textit{PA}(a,B)$.
In Step~\ding{174}, we overwrite node $a$ in array $A$, which has become duplicated,\nadd{ as shown in Fig.~\ref{fig:row_sel}(c)}.
In this step, no copy instruction is needed and the change in CPP number is $-N_\textit{PA}(a,A)$.
In total, 1 copy instruction is needed and $\Delta\#\textit{CPP}=\Delta\#\textit{CPP}_B+N_\textit{PA}(a,B)-N_\textit{PA}(a,A)$.
\nadd{We traverse every node $a$ stored in array $A$ and every other array $B$.}
Among all possible choices, we pick node $a_\textit{best}$ and array $B_\textit{best}$ so that $\Delta\#\textit{CPP}$ is maximized and the row $r_\textit{best}$ containing node $a_\textit{best}$ is our choice.

\subsection{\qdel{Iterative }Improvement Pass}\label{sec:method_iter}
In this section, we \qdel{explain}\qadd{describe} the \ndel{iterative }improvement pass.
As mentioned in Section~\ref{sec:method_over}, \ndel{for each \ndel{run}\nadd{improvement pass}, }we first randomly \qdel{disturb}\qadd{perturb} the \xdel{execution sequence}\xadd{ES} and then apply the \xdel{instruction generation}\xadd{IG} process given the new \xdel{execution sequence}\xadd{ES}.

\begin{wrapfigure}{r}{0.42\linewidth}
  \begin{center}
    \vspace{-2.6em}
    \includegraphics[width=\linewidth]{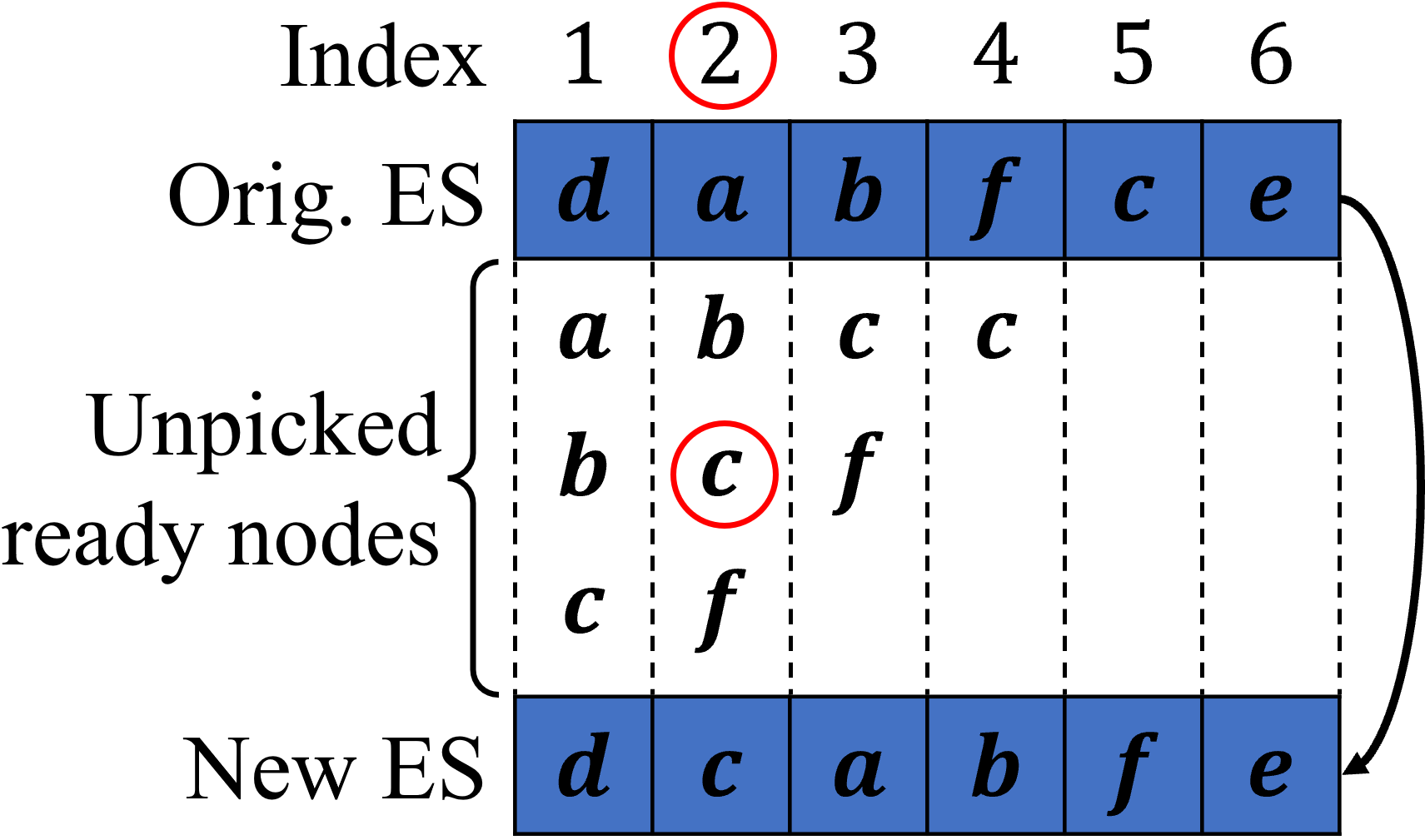}
  \end{center}
  \vspace{-1em}
  \caption{An example of perturbing the execution sequence.}
  \vspace{-0.8em}
  \label{fig:dist}
\end{wrapfigure}
We \qadd{first }describe the \qdel{disturbance process}\qadd{random perturbation} \qdel{with}\qadd{by} an example in Fig.~\ref{fig:dist}\ndel{(a)}.\qadd{ Assume that the original ES is $d,a,b,f,c,e$.}
To \qdel{disturb}\qadd{perturb} \qdel{an \xdel{execution sequence}\xadd{ES}, \textit{e.g.}, $d,a,b,f,c,e$}\qadd{it}, we first randomly pick \qdel{an index}\qadd{a time step index}, \textit{e.g.}, 2.
At \qdel{the }time step \qdel{corresponding to index }2, node $a$ is chosen to be scheduled\qadd{ in the original ES, as shown in Fig.~\ref{fig:dist}\ndel{(a)}}.
However, nodes $b$, $c$, and $f$ are also ready to be computed.
Initially, they are not chosen because \qdel{their priority is worse than $a$}\qadd{they have a lower priority than $a$}, or \qdel{the priority is}\qadd{they have} the same\qadd{ priority as $a$}, but they are not picked due to randomness.
Since picking the node with the highest priority is just a heuristic that cannot\qadd{ always} guarantee the best performance, we attempt to make a different\qadd{ node} choice, hoping to improve the performance.
\qdel{We}\qadd{Thus, we} randomly pick another ready node, \textit{e.g}., node $c$, and \qdel{advance}\qadd{insert} it\qadd{ right before $a$} in the \xdel{execution sequence}\qadd{original }\xadd{ES}\qdel{ to this time step.
In the example}, \qadd{which leads to }the \xdel{execution sequence}\qadd{new }\xadd{ES} \qdel{becomes }$d,c,a,b,f,e$\qadd{ as shown in Fig.~\ref{fig:dist}\ndel{(a)}}.

Given the new ES, we \qadd{propose to apply a}\qdel{apply} \emph{\CombPName{}-driven IG process} to generate the corresponding IS\qdel{ as}\qadd{, which is} shown in \ndel{Fig.~\ref{fig:dist}(b)}the \qdel{last }red box in\qadd{ the bottom-right corner of} Fig.~\ref{fig:overv}.
The \xdel{generation }process is similar to \qdel{that of }the \xdel{\CombPName{}-driven scheduler}scheduling algorithm in Section~\ref{sec:method_sched}.
\qadd{The difference lies in that in each outermost iteration $i$}\qdel{Each time}, instead of \qdel{picking}\qadd{traversing both} the ready \qdel{node $n$}\qadd{nodes} and the \qdel{corresponding array $A$}\qadd{available arrays} \qdel{with}\qadd{to choose the node and the array with} the highest priority, we fix the node \qdel{$n$ according to}\qadd{as} the\qadd{ $i$-th node in the} \xdel{execution sequence}ES and only \qdel{pick}\qadd{traverse} the array\qadd{s to pick the one} with \qadd{the }highest priority.
\qdel{In}\qadd{For example, given the new ES in} \xdel{the example}Fig.~\ref{fig:dist}\ndel{(a)}, we first schedule node $d$ by inserting the \xdel{required }copy instructions needed followed by the computation instruction for node $d$, then move on to node $c$, and so on.

\section{Experimental Results}\label{sec:exp}
This section shows the experimental results.
We implement \xdel{our scheduler}\MethodName{} and \qdel{the}\qadd{some prior state-of-the-art} \qdel{existing }schedulers used for comparison in C++ and perform experiments on a computer with a 24-core 2.4GHz Intel 4214R processor and a 64GB RAM\@ using 12 threads.

Our target is to develop a scheduler that can minimize the number of copy instructions needed for a given array size limit, thus reducing the overall energy consumption.
Hence, the same netlist is given as input to all schedulers.
We use Mockturtle~\cite{EPFLLIB} to synthesize the target functions into the XMGs supported in XMG-GPPIC~\cite{XMG-GPPIC}. 
Since it takes time and energy to compute each node in the netlist in serial, we attempt to minimize the size of the netlist, \textit{i.e.}, the number of nodes, in the synthesis step.
For this purpose, we repeatedly apply the widely-used logic optimization command sequence \textit{resyn2} from ABC~\cite{ABC} to the netlist until its size cannot be reduced any more.\footnote{In our implementation, we use the Mockturtle version of \textit{resyn2}.}

\begin{table}[htbp]
  \centering
  \tabcolsep=2pt
  \caption{Benchmarks used in the experiment.}
  \label{tab:setup}
    \begin{tabular}{cccccccc}
    \toprule
    Benchmark & \#Nodes & PI    & PO    & Benchmark & \#Nodes & PI    & PO \\
    \midrule
    int2float & 199   & 11    & 7     & max   & 1824  & 512   & 130 \\
    router & 201   & 60    & 3     & sin   & 3482  & 24    & 25 \\
    cavlc & 600   & 10    & 11    & sqrt  & 9240  & 128   & 64 \\
    priority & 551   & 128   & 8     & multiplier & 14176 & 128   & 128 \\
    dec   & 304   & 8     & 256   & div   & 12536 & 128   & 128 \\
    adder & 256   & 129   & 256   & log2  & 19760 & 32    & 32 \\
    \bottomrule
    \end{tabular}
\end{table}

The benchmarks used in our experiments are from the EPFL benchmark suite~\cite{EPFLBENCH}.
The information of the optimized netlist of each benchmark is \qdel{shown}\qadd{listed} in Table~\ref{tab:setup}.
For some benchmarks, some of the\qadd{ir} POs are just the PIs or constant $0$ or $1$, so we do not include these POs into the netlists.

Balancing runtime and performance, we set the \qdel{number of runs in our \xdel{scheduler}\xadd{scheduling algorithm} and iterative improvement \xadd{process }to be $R=500$}\qadd{parameter $R$ in Fig.~\ref{fig:overv} as $500$}.
We use the same set of rules as described in Section~\ref{sec:back_rule} for all schedulers.
For existing schedulers that do not allow the output of a node to overwrite one of its inputs, we adjust them slightly to permit such overwriting while still maintaining the core algorithms intact.
The existing schedulers such as GPPIC~\cite{XMG-GPPIC}, SIMPLER~\cite{SIMPLER}, and STAR~\cite{STAR} are relatively fast.
For a fair comparison, we modify them to enhance their performance.
Specifically, we add randomness to their methods.
When there is a tie in priority, we pick a random node among the ones with the highest priority instead of the one with the smallest index.
The improved existing schedulers are run for multiple times until they reach the same runtime as ours, and the best results are reported for each method.
Note that to ensure performance improvement, we run the original scheduling algorithm as the first run in the improved version.
The \xdel{instruction sequence generation}IG strategy used for the previous schedulers is the greedy strategy proposed in~\cite{PREV} since it gives better performance than the naive method proposed in~\cite{XMG-GPPIC}.


We set a different array size limit for each benchmark according to the size of the netlist\ndel{ and the number of rows needed as shown in Table~\ref{tab:copy}} so that \MethodName{} uses 2-4 arrays to implement the benchmark.
The array size \qdel{in terms of the number of rows }and the number of arrays used \qadd{by \MethodName{} }for each benchmark are listed in columns 2 and 3 of Table~\ref{tab:copy}, respectively.
The correctness of the scheduling results is verified by a simple algorithm.
Next, in Section~\ref{sec:exp_copy}, we compare the number of copy instructions obtained by different schedulers.
In Section~\ref{sec:exp_energy}, we compare the resulting energy consumption.
In Section~\ref{sec:exp_ab}, we perform the ablation study.
In Section~\ref{sec:exp_dse}, we \qdel{perform a simple design space exploration on array size}\qadd{study the performance of various schedulers when different array sizes are considered}.
\subsection{Performance on Reducing Copy Instruction Number}\label{sec:exp_copy}
\begin{table*}
  \centering
  \caption{\qdel{\#Copy}\qadd{Numbers of copy} instructions obtained by various schedulers\qadd{ and their runtimes}.}
\scriptsize
\begin{tabular}{cccccccccccccc}
    \toprule
    \multirow{2}[4]{*}{Benchmark} & \multirow{2}[4]{*}{ArraySize} & 
    \multirow{2}[4]{*}{\#Array} & \multicolumn{8}{c}{\#Copy}                                    & \multicolumn{3}{c}{Time(s)} \\
\cmidrule(r){4-11}\cmidrule(r){12-14}   &       &       & \cite{XMG-GPPIC} & \cite{SIMPLER} & \cite{STAR}  & \cite{PREV} & \textbf{\MethodName{}}  & \cite{PREV}+IG & NoImpr & \cite{PREV}+Impr & \cite{PREV} & \cite{XMG-GPPIC,SIMPLER,STAR}, \MethodName{} & NoImpr \\
    \toprule
    int2float & 16 &2   & 252   & 220   & 222   & \underline{211} & \textbf{87}    & 105   & 114   & 179   & 43    & 6     & 3 \\
    router & 64  &2  & 188   & 170   & \underline{169} & 176   & \textbf{70}    & 88    & 71    & 152   & 19    & 9     & 7 \\
    cavlc & 64  &2  & 421   & 205   & 281   & \underline{54} & \textbf{19}    & 24    & 26    & 49    & 227   & 72    & 41 \\
    priority & 128  &2 & \underline{238} & \underline{238} & \underline{238} & \underline{238} & \textbf{128}   & 128   & 128   & 238   & 235   & 30    & 20 \\
    dec   & 256 &2  & \underline{18} & 20    & \underline{18} & \underline{18} & \textbf{9}     & 9     & 9     & 18    & 92    & 24    & 15 \\
    adder & 256 &2  & \underline{512} & \underline{512} & \underline{512} & \underline{512} & \textbf{256}   & 256   & 256   & 512   & 42    & 29    & 21 \\
    max   & 256 &4  & 2294  & 1726  & 1725  & \underline{1723} & \textbf{937}   & 1250  & 1156  & 1723  & 149   & 97    & 71 \\
    sin   & 256 &2  & 687   & 1574  & 572   & \underline{423} & \textbf{120}   & 132   & 136   & 361   & 809   & 502   & 483 \\
    sqrt  & 256 &3  & 4915  & 4639  & 4701  & \underline{4331} & \textbf{1290}  & 1307  & 1478  & 4039  & 1437  & 135   & 111 \\
    multiplier & 256 &2  & 16009 & 7303  & 11231 & \underline{5961} & \textbf{1440}  & 1736  & 1643  & 5525  & 1131  & 708   & 596 \\
    div   & 256 &3  & 6144  & 6013  & 7609  & \underline{5812} & \textbf{872}   & 926   & 965   & 4986  & 1339  & 700   & 439 \\
    log2  & 256 &4  & 11388 & 14702 & 12581 & \underline{7043} & \textbf{3311}  & 3878  & 3793  & 6414  & 2465  & 886   & 732 \\
    GEOMEAN &    &   & 922.4  & 854.7  & 835.3  & 622.5  & \textbf{228.8}  & 258.2  & 257.9  & 574.7  & 262.9  & 88.6  & 63.1  \\
    \bottomrule
    \end{tabular}%
  \label{tab:copy}%
\end{table*}%

In Table~\ref{tab:copy}{, columns 4--8 \qdel{compare}\qadd{list} the numbers of copy instructions obtained by the existing schedulers~\cite{XMG-GPPIC,SIMPLER,STAR,PREV}\qadd{,} and \xdel{ours}\MethodName{}, \qdel{and}\qadd{while} columns 12 and 13 \qdel{compare}\qadd{list} their runtime\qadd{s}.
%
For each benchmark, we highlight the best scheduling result, \textit{i.e.}, the one with the least number of copy instructions, in bold.
We can see that \MethodName{}\ndel{ marked in bold in the table} achieves\xdel{ the same or} \qdel{better}\qadd{the fewest copy instructions among all methods} \qdel{result }for all benchmarks\xdel{ except \textit{int2float}}.
Compared to XMG-GPPIC~\cite{XMG-GPPIC}, SIMPLER~\cite{SIMPLER} and STAR~\cite{STAR}, \xdel{our scheduler}\qdel{\MethodName{}}\qadd{it} reduces the number of copy instructions by 75.2\%, 73.2\%, and 72.6\%, respectively, on average using the same runtime.
Compared to the recent work~\cite{PREV}, \xdel{our scheduler}\qdel{\MethodName{}}\qadd{it} reduces the number of copy instructions by 63.2\% on average with only 33.7\% of the runtime.
Compared to the best result obtained by\qadd{ the} existing schedulers~\cite{XMG-GPPIC,SIMPLER,STAR,PREV} for each benchmark\qdel{ that}\qadd{, which} is \ndel{marked in bold}underlined in the table, \MethodName{} still reduces the number of copy instructions by 63.1\% on average.
For \qdel{the frequently}\qadd{two commonly}-used large benchmarks\qdel{ such as}\qadd{,} multiplier and divisor, \MethodName{} only needs less than 1/4 and 1/6 copy instructions, respectively, compared to the best result.
The results demonstrate the ability of \MethodName{} \qdel{to reduce}\qadd{in reducing} the number of copy instructions.
\vspace{-0.1em}
\subsection{Performance on Reducing Energy Consumption}\label{sec:exp_energy}
\vspace{-0.5em}
\begin{figure}[!htbp]
    \centering
    \includegraphics[scale=0.65]{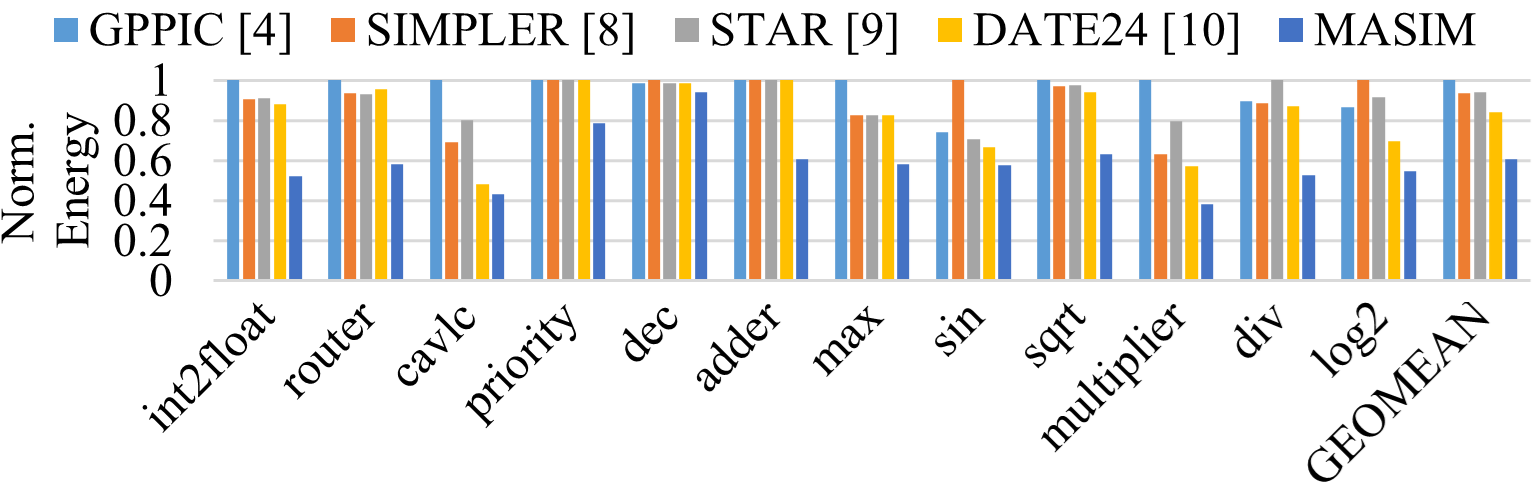}
    \caption{Energy comparison.}
    \label{fig:energy}
\end{figure}
\vspace{-0.5em}
In this section, we compare the energy consumption of the \xdel{instruction sequences}ISs obtained by different schedulers.
To obtain the energy consumption of a given \xdel{instruction sequence}IS, we accumulate the energy consumption of each instruction\ndel{, \textit{i.e.}, computation instruction and copy instruction}.
The energy consumption of each \qdel{kind}\qadd{type} of instruction, \textit{i.e.}, computation instruction and copy instruction, for each array size is obtained using the method from~\cite{XMG-GPPIC}.
For each benchmark, we normalize the energy to the maximum of the five methods.
As shown in Fig.~\ref{fig:energy}, compared to XMG-GPPIC~\cite{XMG-GPPIC}, SIMPLER~\cite{SIMPLER}, STAR~\cite{STAR}, and DATE24~\cite{PREV}, \MethodName{} reduces the total energy consumption by 39.4\%, 35.4\%, 35.7\%, and 28.0\%, respectively\qadd{, on average}.
Note that the energy reduction is less prominent than the copy instruction number reduction, since some portion of the energy is spent on computation instructions that cannot be saved.
The results show the ability of \MethodName{} \qdel{to reduce}\qadd{in reducing} energy consumption.
\subsection{Ablation Study}\label{sec:exp_ab}
There are \ndel{three main components}two main parts in our schedulers: \ndel{\CombPName{}-driven \xdel{instruction sequence generator (IG)}IG\qadd{ process} that generates the \xdel{instruction sequence}IS given the \xdel{execution sequence}ES of the nodes, }\CombPName{}-driven \xdel{scheduler}scheduling algorithm that can \ndel{directly }determine the \xdel{instruction sequence}IS\ndel{,} and the iterative improvement process that improves the obtained IS.
In the iterative improvement process, the \CombPName{}-driven IG strategy \qdel{illustrated}\qadd{described} in Section~\ref{sec:method_iter}\qdel{ that}\qadd{, which} generates the IS given the ES\qadd{,} is also an important component.
We study the effect of each of them as follows.

\subsubsection{\CombPName{}-Driven IG}
In Table~\ref{tab:copy}, the column \mbox{``\hspace{1sp}\cite{PREV}''} lists the number of copy instructions obtained using the \xdel{execution sequence}ES given by the scheduler in~\cite{PREV} and the greedy IG~\cite{PREV}.
The column \mbox{``\hspace{1sp}\cite{PREV}+IG''} lists the number of copy instructions obtained using the same \xdel{execution sequence}ES and our \CombPName{}-driven IG.
The results show that our \CombPName{}-driven IG achieves fewer copy instructions for all benchmarks\xdel{ except \textit{int2float}}.
On average, it reduces the number of copy instructions by 58.5\%, showing the strength of our \CombPName{}-driven IG.
Note that both IG methods are very fast compared to the scheduling algorithms, so we do not compare their runtime\qadd{s}.

\subsubsection{\CombPName{}-Driven Scheduling Algorithm}
In Table~\ref{tab:copy}, the column \mbox{``NoImpr''} lists the number of copy instructions obtained using our \CombPName{}-driven \xdel{scheduler}scheduling algorithm without the following iterative improvement process, and its runtime is listed in the last column.
Compared to the best existing scheduler~\cite{PREV}, it reduces the number of copy instructions by 58.6\%.
Even compared to the enhanced version of~\cite{PREV} using the \CombPName{}-driven IG, \textit{i.e.}, \mbox{``\hspace{1sp}\cite{PREV}+IG''}, it achieves a bit fewer copy instructions by only using 24\% of the runtime, showing the efficiency of our \CombPName{}-driven scheduling algorithm.

\subsubsection{Iterative Improvement Process}
In Table~\ref{tab:copy}, the column \mbox{``\hspace{1sp}\cite{PREV}+Impr''} lists the number of copy instructions obtained based on the result of~\cite{PREV} and improved using our proposed iterative improvement process.
Note that to make the comparison fair, the IG used in this version is the greedy IG used in~\cite{PREV}.
Comparing \mbox{``\hspace{1sp}\cite{PREV}+Impr''} and \mbox{``\hspace{1sp}\cite{PREV}''}, the iterative improvement process reduces the number of copy instructions by 7.7\%.
Comparing \mbox{\MethodName{}} and \mbox{``NoImpr''}, which are \xdel{our scheduler\qadd{s}}the \MethodName{s} with and without\qadd{ the} iterative improvement process, respectively, the iterative improvement process reduces the number of copy instructions by 11.3\%.
This shows the strength of our iterative improvement process.

\subsection{\qdel{Design Space Exploration}\qadd{The Impact of Different Array Sizes}}\label{sec:exp_dse}

\begin{wrapfigure}{r}{0.5\linewidth}
  \begin{center}
    \vspace{-2em}
    \includegraphics[width=\linewidth]{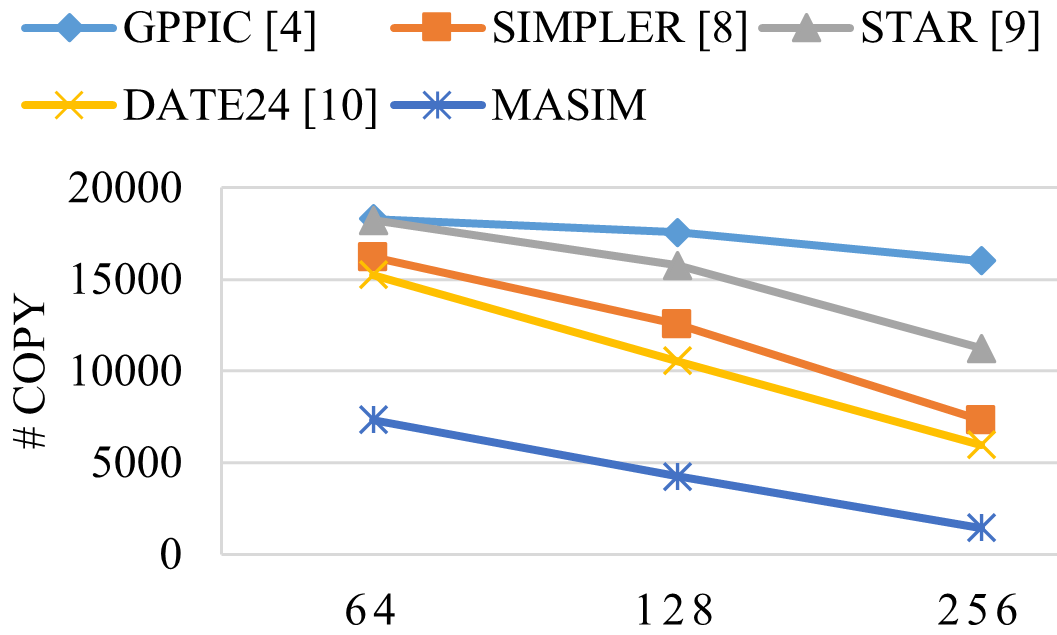}
  \end{center}
  \vspace{-1em}
  \caption{Number of copy instructions needed for \textit{multiplier} using different array sizes.}
  \vspace{-1em}
  \label{fig:dse}
\end{wrapfigure}
In this section, we \qdel{perform a design space exploration on the number of copy instructions needed to implement the benchmark \textit{multiplier} on arrays of different sizes}\qadd{study the impact of different array sizes on the number of copy instructions needed to implement the benchmark \textit{multiplier} for various schedulers}.
\qadd{We consider three array sizes: $64$, $128$, and $256$.}
The result is shown in Fig.~\ref{fig:dse}.
As expected, \qdel{when the array size decreases, }the number of copy instructions needed \qdel{increases greatly}\qadd{decreases with the array size,} and \MethodName{} outperforms \qadd{the }existing methods for all array sizes.

\section{Conclusion}\label{sec:conc}
\qdel{Copy instructions consume a lot of energy when using SIMD IMC hardware to calculate a target function\qdel{.
However} \qdel{such copy instructions}\qadd{but they} cannot be avoided when implementing complex functions on memory arrays of limited sizes.
The existing schedulers mainly focus on reducing the number of needed rows so that functions can be \xdel{computed}\qdel{calculated}\qadd{implemented} within a single array and do not \qdel{put}\qadd{spend} much effort \qdel{on}\qadd{in} reducing the number of copy instructions needed when using multiple arrays.}%
In this work, we propose\qadd{ \MethodName{},} \qadd{a multi-array scheduler for SIMD IMC, which can significantly reduce the number of energy-consuming copy instructions needed when multiple arrays are used in SIMD IMC. It involves a priority}\qdel{\CombPName{}}-driven scheduling algorithm \qdel{using}\qadd{that uses} the number of copy instructions as the primary priority and the \xdel{change in}number of \PName{s}\qadd{, which is a new metric we propose,} as the secondary priority.
To further reduce \qadd{the }copy instruction number, we propose an iterative improvement process.
\MethodName{} reduces the number of copy instructions by 63.2\% on average compared to the best \qdel{state-of-the-art}\qadd{prior} scheduler, leading to a 28.0\% reduction in energy consumption.
In this work, we assume that the energy consumption of \qdel{the}\qadd{a} copy instruction\qdel{s} is a constant.
However, it generally takes less energy to copy data between arrays that are located close to each other on the chip.
We will \qdel{take this into account}\qadd{extend \MethodName{} to handle this situation} in \qdel{our}\qadd{the} future\qdel{ work}.
\newpage
\bibliographystyle{IEEEtran}
\bibliography{cite}
\end{document}